\documentclass[letterpaper]{jpconf}
\usepackage{graphicx}

\begin{document}

\title{Differences and analogies between quantum chromodynamics and ferromagnets}

\author{C P Hofmann$^1$}

\address{$^1$ Facultad de Ciencias, Universidad de Colima, Bernal D\'iaz del Castillo 340, Colima C.P.\ 28045, Mexico}

\ead{christoph.peter.hofmann@gmail.com}

\begin{abstract}
The low-energy physics of quantum chromodynamics (QCD) and ferromagnets is dominated by Goldstone bosons. While the effective theory of
QCD - chiral perturbation theory - is well established in the particle physics community, the systematic studies of ferromagnetic systems
within the effective Lagrangian framework are not well-known. We analyze the low-temperature properties of ferromagnets in one, two and
three space dimensions up to three-loop order in the effective expansion, i.e., beyond the accuracy of any previous results obtained with
conventional condensed matter methods. In particular, in the nonrelativistic domain, the effective method perfectly works in one space
dimension.
\end{abstract}

\section{Motivation}

The effective field theory of quantum chromodynamics (QCD), chiral perturbation theory, is very-well established in particle physics. The
method was originally devised for zero temperature \cite{Wei79,GL84,GL85}, but has been extended to finite temperature soon after
\cite{GL87a,GL87b}. There are many excellent outlines of the method available -- the interested reader may want to consult
Refs.~\cite{Leu95,Sch03,Goi04,Bur07,Brau10}. In the present overview, we are interested in the description of relativistic and
nonrelativistic systems at nonzero temperature. In fact, the low-temperature properties of QCD have been systematically analyzed in a
series of papers a long time ago \cite{GL87a,GL87b,GL89}. One important result of these studies is the three-loop formula for the order
parameter, i.e., the quark condensate, which takes the following form \cite{GL89},
\begin{equation}
\frac{\langle \bar{q} q \rangle(T,m_q)}{\langle 0 | \bar{q} q | 0 \rangle}
= 1 - {\beta}_0 T^2 - {\bf {\beta}_1 T^4} - {\bf {\beta}_2 T^6 \, ln T} + {\cal O}({\bf T^8}) ,
\end{equation}
where $\langle 0 | \bar{q} q | 0 \rangle$ is the quark condensate at zero temperature. Note that the boldfaced terms are those
contributions that result from the interaction among the Goldstone bosons, while the leading temperature-dependent term corresponds to
free Goldstone bosons.

The effective field theory method, however, is not restricted to the relativistic domain. Indeed, the effective Lagrangian technique has
been transferred to nonrelativistic systems in Ref.~\cite{Leu94a}. In particular, the effective Lagrangian for the ferromagnet was
established in that reference. In analogy to the low-temperature expansion of the quark condensate and other thermodynamic quantities, in
this talk, we will be interested in the behavior of the (spontaneous) magnetization and other observables of ferromagnets in one, two and
three space dimensions -- detailed information can be found in the original papers \cite{Hof02,Hof11a,Hof12a,Hof12b,Hof13a,Hof13b}. We
will present a three-loop analysis of the partition function in each case and point our both differences and analogies between QCD and
ferromagnets. One of our main themes is to determine at what order in the low-temperature expansion, the interaction among the Goldstone
bosons (the spin waves or magnons in the case of the ferromagnet, the pseudoscalar mesons in the case of QCD) does manifest itself.

\section{From the underlying theory to the effective field theory}

\subsection{Symmetries, Goldstone theorem and order parameters}

In the chiral limit, i.e. when the light quark masses $m_u, m_d, m_s$ are set to zero, the QCD Lagrangian,
\begin{equation}
{\cal L}_{\mbox{\scriptsize{QCD}}} = - \mbox{$\frac{1}{2g^2}$} \mbox{tr}_c G_{\mu \nu} G^{\mu \nu}
+ {\bar q} i \gamma^{\mu} D_{\mu} q - {\bar q} m q ,
\end{equation}
is invariant under the chiral transformation {SU(3)}$_R \times$ {SU(3)}$_L$. The QCD vacuum, on the other hand, is only invariant under
{SU(3)}$_V$: the chiral symmetry is spontaneously broken, and the quark condensate serves as an order parameter. Goldstone's theorem then 
predicts eight pseudoscalar mesons in the low-energy spectrum of QCD: three pions, four kaons and the $\eta$-particle. In the real world
we rather have $m_u, m_d, m_s \neq 0$, i.e., the chiral symmetry is only approximate, such that there emerge pseudo-Goldstone bosons: the
pseudoscalar mesons are not massless, but their masses are still small compared to the mass of, e.g., the proton.

An analogous situation arises in the context of ferro- and antiferromagnets that are described by the Heisenberg model on the underlying
or microscopic level,
\begin{equation}
{\cal H}_0 = - J \sum_{n.n.} {\vec S}_m \! \cdot {\vec S}_n , \qquad \quad J = const.
\end{equation}
Depending on the sign of the exchange integral $J$, a ferromagnetic ($J > 0$) or antiferromagnetic ($J < 0$) ground state is preferred.
But again, whereas the Hamiltonian ${\cal H}_0$ is invariant under the group G = O(3), describing rotations in the space of the spin
variables, the ground state is only invariant under the subgroup H = O(2), describing rotations around the third axis. In the case of the
three-dimensional ferromagnet, the spontaneous magnetization $\Sigma$ serves as an order parameter. Below the Curie temperature, we are in
the broken phase ($\Sigma \neq 0$), above the Curie temperature, the system is in the symmetric phase ($\Sigma = 0$). The Curie
temperature is thus the analog of the chiral phase transition temperature in QCD.

Goldstone's theorem for nonrelativistic systems \cite{Lan66,GHK68,CN76} also implies that there are low-energy excitations in the spectrum
whose energy tends to zero for large wavelengths: $\omega \to 0$ for $|{\vec k}| \to 0$. However, in the case of the ferromagnet, although
we have two magnon fields, we only have one magnon particle that follows a quadratic dispersion relation \cite{Leu94a,Hof99a}. These
magnons or spin waves can be viewed as collective fluctuations of the spins.

The Heisenberg model can be extended by incorporating an external magnetic field $\vec H$ that couples to the total spin vector,
\begin{equation}
{\cal H} = - J \sum_{n.n.} {\vec S}_m \! \cdot {\vec S}_n -
{\mu}{\sum_n}{\vec S}_n \! \cdot {\vec H} = {\cal H}_0 + {\cal H}_{sb} .
\end{equation}
Much like the quark masses in the QCD Lagrangian explicitly break chiral symmetry, the magnetic field is a symmetry breaking parameter
that explicitly breaks the spin rotation symmetry O(3) of the Heisenberg Hamiltonian. For small quark masses and for weak magnetic fields,
these quantities represent perturbations of the spontaneously broken symmetry.

\subsection{Relativistic and nonrelativistic effective Lagrangians}

The construction of effective Lagrangians is based on an analysis of the symmetries in the underlying theory (QCD Lagrangian, Heisenberg
Hamiltonian), in particular of the spontaneously broken symmetry \cite{Wei79,CWZ69,CCWZ69,Leu94b}. The degrees of freedom in the effective
Lagrangian are the Goldstone bosons that dominate the low-energy properties of the system. One then proceeds to write down the most
general possible Lagrangian that includes all terms consistent with the symmetries identified in the underlying theory, and then
calculates observables with this Lagrangian to the desired order of perturbation theory. This corresponds to an expansion of observables
in powers of momenta (or temperature), i.e., to a derivative expansion of the effective Lagrangian. In chiral perturbation theory,
Lorentz-invariance implies that the derivative expansion consists of a series 
\begin{equation}
{\cal L}_{eff} = {\cal L}_{eff} (\pi, \partial \pi, \partial ^2 \pi,
\ldots, m) = {\cal L}_{eff}^2 + {\cal L}_{eff}^4 + {\cal L}_{eff}^6+ \ldots ,
\end{equation}
where the leading term (order $p^2$) is given by
\begin{equation}
{\cal L}^2_{eff} = \mbox{$\frac{1}{4}$}F^2_{\pi} \, \mbox{tr} \, \{ \partial_\mu U \partial^\mu
U^{\dagger} + \chi (U +U^{\dagger}) \}, \qquad \chi = 2 B m .
\end{equation}
Note that the Goldstone boson fields, i.e. the pseudoscalar mesons, are collected in the matrix $\pi(x)$ that is related to $U(x)$ by
\begin{equation}
U  =  \exp[2 i \pi(x) / F_{\pi}] .
\end{equation}
At leading order ${\cal O}(p^2)$, we have two effective constants: $F_{\pi}$ (pion decay constant) and $B$.

The next-to-leading order (${\cal O}(p^4))$ chiral effective Lagrangian,
\begin{equation}
\label{LeffCHPTp4}
{\cal L}_{eff}^4 = L_1  \, \mbox{tr} \, \{ D_\mu U^\dagger D^\mu U \}^2 + \dots 
+ L_5  \, \mbox{tr} \, \{ D_\mu U^\dagger D^\mu U (\chi^\dagger U + U^\dagger \chi) \} + \dots ,
\end{equation}
already involves ten effective constants ($L_1, \dots$) that parametrize the physics of the underlying theory (QCD). Note that these
effective constants have to be determined by experiment (e.g. pion-pion scattering) or in numerical simulations -- symmetry does not fix
their actual values. It is important to point out, however, that symmetry does unambiguously determine the derivative structure of the
various terms in the effective Lagrangian.

In the nonrelativistic domain, charge densities are allowed to pick up nonzero expectation values, since Lorentz-invariance no longer
represents a symmetry of the system. This is precisely what happens in the case of the ferromagnet, where the spontaneous magnetization
$\Sigma$ shows up as a leading-order effective constant of a term that involves one time derivative only and thus dominates the low-energy
dynamics \cite{Leu94a},
\begin{equation}
{\cal L}_{eff}^{F} = \Sigma \frac{{\partial}_0 U^1U^2 - {\partial}_0
U^2U^1}{1 + U^3} + {\Sigma}H^i U^i - \mbox{$ \frac{1}{2}$} F^2 D_rU^iD_rU^i .
\end{equation}
Note that the spontaneous magnetization also occurs in the term that involves the magnetic field ${\vec H} = (0,0,H)$. The spin waves
correspond to the fluctuations of the magnetization vector $\vec U$, defined by
\begin{equation}
\vec{U} = (U^1,U^2,U^3) = ({\pi}^1, {\pi}^2, \sqrt{1- {\pi}^a {\pi}^a}), \quad a=1,2 \, , \qquad |\vec{U}| = 1 .
\end{equation}
Note the difference with respect to chiral perturbation theory, where the leading-order effective Lagrangian reads
\begin{equation}
{\cal L}^{QCD}_{eff} = \mbox{$\frac{1}{4}$}F^2_{\pi} \,\mbox{tr}\, \{ \partial_\mu U \partial^\mu
U^{\dagger} + 2 B m (U +U^{\dagger}) \},  \quad U  =  \exp[2 i \pi(x) / F_{\pi}] .
\end{equation}
Accordingly, the dispersion relation of pseudoscalar mesons takes a linear (relativistic) form
\begin{equation}
\omega = \sqrt{c^2 {\vec k}^2 + c^4 M^2} , \qquad M^2 = (m_u + m_d) B ,
\end{equation}
while for ferromagnetic magnons it is quadratic,
\begin{equation}
\omega = \gamma {\vec k}^2 + \mu H, \qquad \gamma = F^2 / \Sigma .
\end{equation}
The constant $F$ in the leading-order effective Lagrangian of the ferromagnet is thus related to the helicity modulus
$\gamma = F^2 / \Sigma$.

\begin{figure}
\begin{center}
\includegraphics[width=15.5cm]{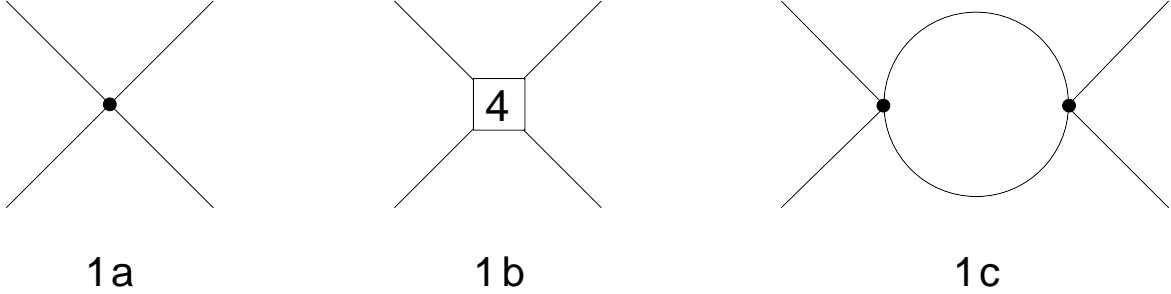}
\end{center}
\caption{\it Loops in chiral perturbation theory are suppressed by two powers of momentum. The ultraviolet divergences of the one-loop
graph 1c can be absorbed into next-to-leading effective constants contained in the tree graph 1b.}
\label{figure0}
\end{figure}

An essential point for the effective framework to be consistent is that the ultraviolet divergences that occur in loop graphs can
systematically be absorbed by effective constants, order by order in the low-energy expansion. In chiral perturbation theory, the
pion-pion scattering graphs up to next-to-leading order are shown in Fig.~\ref{figure0}.
While the tree graph 1a of order $p^2$ is finite, the one-loop graph 1c of order $p^4$ is singular in the ultraviolet. However, these
divergences can be absorbed into coupling constants of order $p^4$ showing up in graph 1b. The crucial point is that these two graphs do
"communicate", because Goldstone-boson loops in 3+1 dimensions are suppressed by two powers of momentum: graphs 1b and 1c are of the same
momentum order $p^4$.

It is important to point out that the suppression of loops depends both on the dispersion relation of the system and on the space
dimension $d_s$. Consider a diagram with a single Goldstone boson loop. The corresponding integrals are 
\begin{equation}
\int \! \frac{{\mbox{d}}E \, {\mbox{d}}^{d_s} p}{E^2 - {\vec p}^2} \ \propto \ {\cal P}^{d_s-1} , \qquad 
\int \! \frac{{\mbox{d}}E \, {\mbox{d}}^{d_s} p}{E - {\vec p}^2} \ \propto \ {\cal P}^{d_s} .
\end{equation}
In a Lorentz-invariant framework we thus have
\begin{itemize}
\item{$d$=3+1: Loops are suppressed by two powers of momentum}
\item{$d$=2+1: Loops are suppressed by one power of momentum}
\end{itemize}
On the other hand, in a nonrelativistic setting, like for the ferromagnet that exhibits a quadratic dispersion relation, we conclude
\begin{itemize}
\item{$d$=3+1: Loops are suppressed by three powers of momentum}
\item{$d$=2+1: Loops are suppressed by two powers of momentum}
\item{$d$=1+1: Loops are suppressed by one power of momentum}
\end{itemize}
Hence there is an analogy between ferromagnets in two space dimensions and QCD: in the effective field theory expansion, loops are
suppressed by two powers of momentum in either case. Accordingly, the structure of the low-temperature series, as we discuss in the next
section, exhibit common features. But there is also an important difference: whereas the effective loop expansion fails in 1+1 space
dimensions in the case of Lorentz-invariant theories, it is perfectly consistent in the nonrelativistic domain: in particular,
ferromagnetic spin-chains are accessible by effective field theory.

\section{Low-temperature expansion of the partition function}

\subsection{Analogies between QCD and ferromagnets in two space dimensions}

\begin{figure}
\begin{center}
\includegraphics[width=16cm]{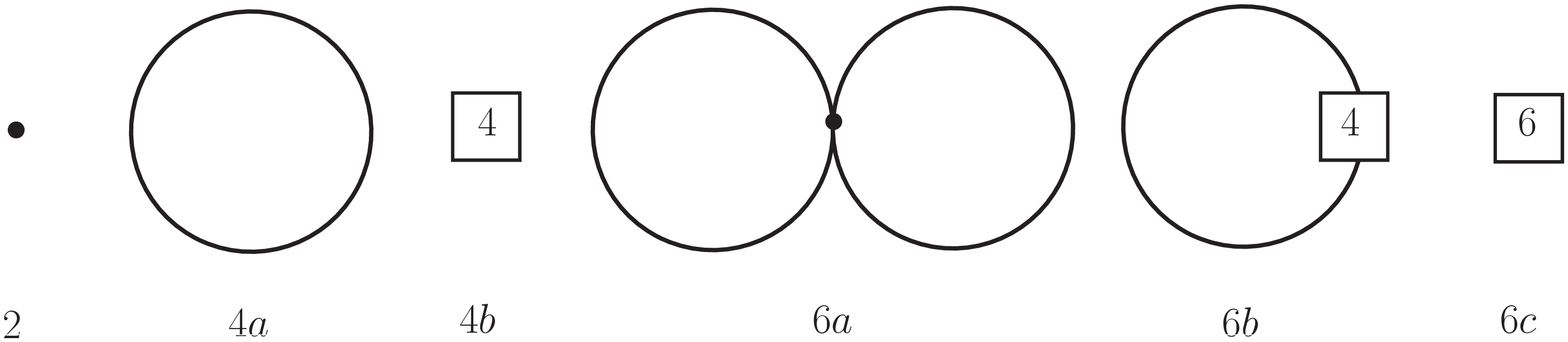}

\vspace{4mm}

\includegraphics[width=16cm]{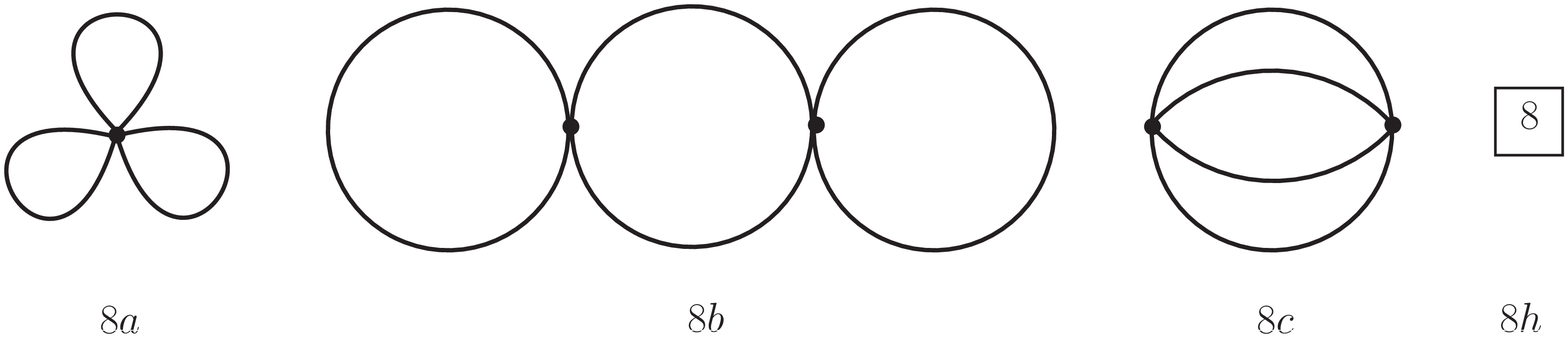}

\vspace{4mm}

\includegraphics[width=16cm]{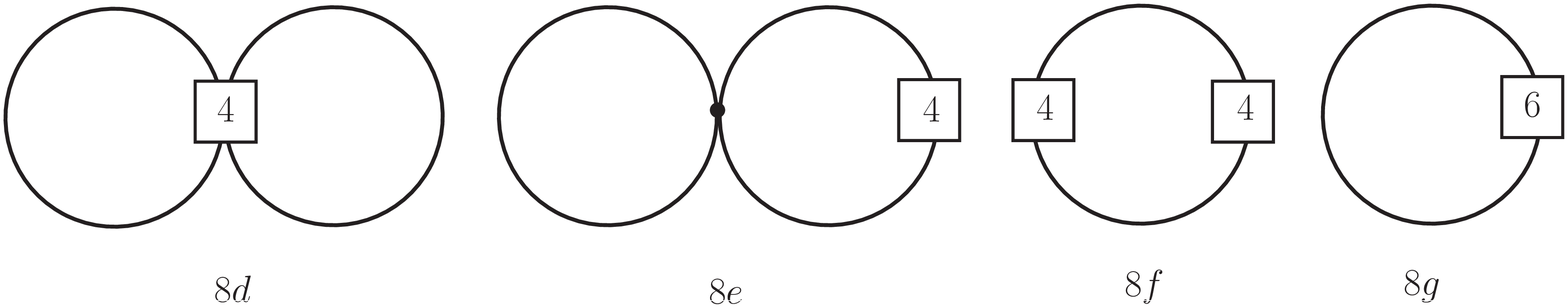}
\end{center}
\caption{\it Feynman graphs: QCD partition function and partition function of the $d_s$=2 ferromagnet up to three-loop order $p^8$ in the
perturbative expansion. Note that loop graphs are suppressed by two powers of momentum. In the case of the ferromagnet, the
$T$-independent graphs 4b, 6c, 8h are absent \cite{Hof02}.}
\label{figure1}
\end{figure}

The relevant Feynman graphs corresponding to the low-temperature expansion of the partition function, both for QCD or ferromagnets in
two space dimensions, are depicted in Fig.~\ref{figure1}.
While the next-to-leading order chiral effective Lagrangian ${\cal L}^4_{eff}$ was given in Eq.~(\ref{LeffCHPTp4}), the subleading pieces
in the effective Lagrangian for the ferromagnet are
\begin{eqnarray}
{\cal L}^2_{eff} & = & \Sigma  \frac{{\varepsilon}_{ab} {\partial}_0 U^a
U^b}{1 + U^3} \; + \; {\Sigma} \mu H U^3 \; - \; \mbox{$ \frac{1}{2}$} F^2
{\partial}_rU^i {\partial}_rU^i , \nonumber \\
{\cal L}^4_{eff} & = & l_1 ({\partial}_r U^i {\partial}_r U^i)^2 + l_2
({\partial}_r U^i {\partial}_s U^i)^2 + l_3 \Delta U^i \Delta U^i , \nonumber \\
{\cal L}^6_{eff} & = & c_1 U^i {\Delta}^3 U^i , \nonumber \\
{\cal L}^8_{eff} & = & d_1 U^i {\Delta}^4 U^i .
\end{eqnarray}
Note that the higher-order pieces ${\cal L}^6_{eff}$ and ${\cal L}^8_{eff}$ are irrelevant for the Goldstone boson interaction according to
Fig.~\ref{figure1}: they only occur in one-loop or tree graphs. Note that the relation between momentum and temperature depends on the
dispersion relation: in chiral perturbation theory we have $p \propto E \propto T$, whereas for the ferromagnet it reads
$p^2 \propto E \propto T$. 

The evaluation of the partition function for ferromagnets in two space dimensions and QCD is discussed in detail in
Refs.~\cite{GL89,Hof12a,Hof12b}. Here we quote the results for the pressure,
\begin{equation}
P_{\mbox{\scriptsize{QCD}}} = b_0 T^4 + {\bf b_1 T^6} + {\bf b_2 \, T^8 \, ln T} + {\cal O}({\bf T^{10}}) ,
\end{equation}

\begin{equation}
P_{Ferro2D} = {\hat a}_0 T^2 + {\hat a}_1 T^3 + {\bf {\hat a}^A_2 \, T^4} + {\bf {\hat a}^B_2 \, T^4 \, ln T} + {\cal O}({\bf T^5}) ,
\end{equation}
and the order parameter: while the low-temperature expansion of the quark condensate is
\begin{equation}
\frac{\langle \bar{q} q \rangle(T,m_q)}{\langle 0 | \bar{q} q | 0 \rangle}
= 1 - {\beta}_0 T^2 - {\bf {\beta}_1 T^4} - {\bf {\beta}_2 T^6 \, ln T} + {\cal O}({\bf T^8}) ,
\end{equation}
for the magnetization we obtain
\begin{equation}
\frac{\Sigma(T,H)}{{\Sigma}} = 1 - {\hat \alpha}_0 T - {\hat \alpha}_1 T^2
- {\bf {\hat \alpha}^A_2 \, T^3} - {\bf {\hat \alpha}^B_2 \, T^3 \, ln T} + {\cal O}({\bf T^4}) .
\end{equation}
Note that all interaction contributions in the above expressions are boldfaced. Again there is this analogy between QCD and ferromagnets
in two space dimensions: due to the identical loop suppression, the structure of temperature powers is analogous. Also, so-called chiral
logarithms arise in the renormalization of loop graphs, and logarithms in the low-temperature series emerge at three-loop order. More
precisely, as in chiral perturbation theory, the cateye-graph for ferromagnets in two space dimensions is logarithmically divergent, 
\begin{equation}
T^{d_s+2} \, {(\mu H)}^{\frac{d_s-2}{2}} { \Bigg\{ \sum_{n=1}^{\infty} \,
\frac{e^{- \mu H n /T}}{n^{\frac{d_s+2}{2}}} \Bigg\} }^2 \, \Gamma(1-\frac{d_s}{2}) .
\end{equation}
This expression is dimensionally regularized in the space dimension $d_s$. In two space dimensions, the cateye-graph 8c exhibits an
ultraviolet divergence. This singularity can be absorbed into the next-to-leading order low-energy constants $l_1$ and $l_2$ from
${\cal L}^4_{eff}$ contained in the two-loop graphs 8d and 8e. This logarithmic renormalization of the effective constants leads to
chiral logarithms -- $\log(H/\mu)$ in the case of the ferromagnet, and $\log(M/\mu)$ in the case of QCD -- where $\mu$ is the
renormalization scale. Moreover, in the limit ${\vec H} \to 0$ (much like in the chiral limit $m_q \to 0$), the cateye graph gives rise to
logarithmic terms $\propto  T^n \, ln T$ in the low-temperature series for the pressure and the order parameter at the three-loop level,
both for QCD ($n$=8,6) and ferromagnetic films ($n$=4,3).

There is, however, an important difference at the two-loop level as far as the Goldstone boson interaction is concerned. While the
two-loop graph 6a (see Fig.~\ref{figure1}) does contribute in the case of QCD, the same graph turns out to be zero for the ferromagnet
(in any space dimension) because of space rotation invariance or parity. As a consequence, the terms proportional to $T^3$ ($T^2$) in the
low-temperature series for the pressure (magnetization) for the ferromagnet in two space dimensions are exclusively determined by free
magnons. On the other hand, the analogous terms in the low-temperature expansion of the QCD pressure and the quark condensate ($T^6$ and
$T^4$, respectively) are affected by the interaction.

\subsection{Ferromagnets in odd space dimensions are different}

\begin{figure}
\begin{center}
\includegraphics[width=16cm]{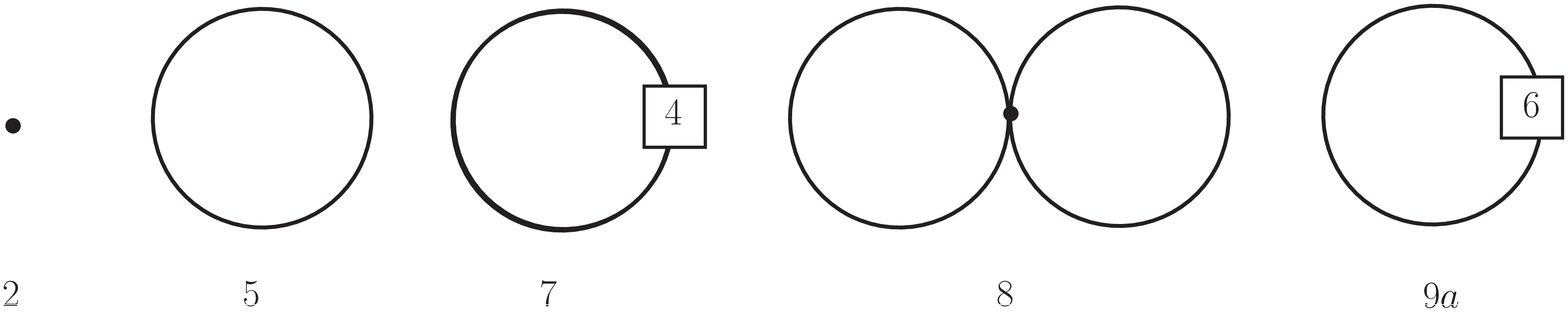}

\vspace{4mm}

\includegraphics[width=16cm]{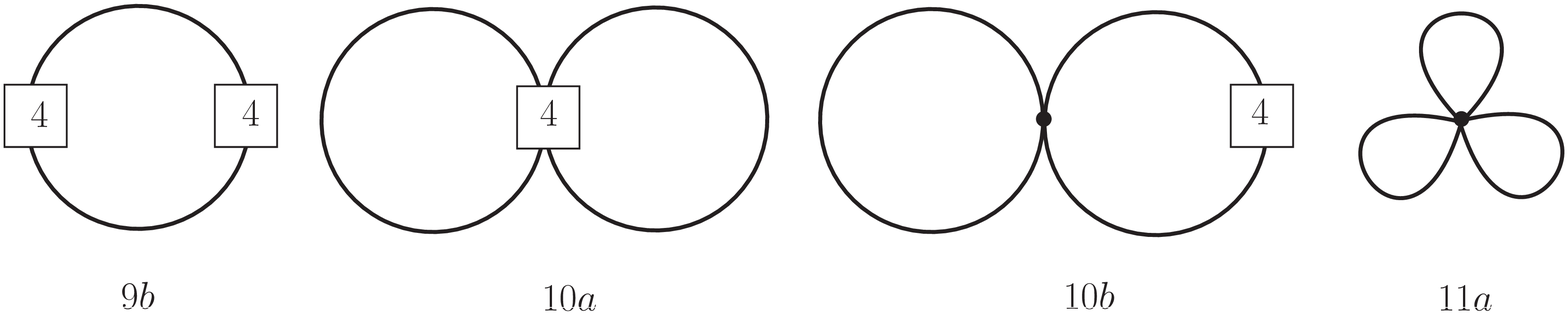}

\vspace{4mm}

\includegraphics[width=16cm]{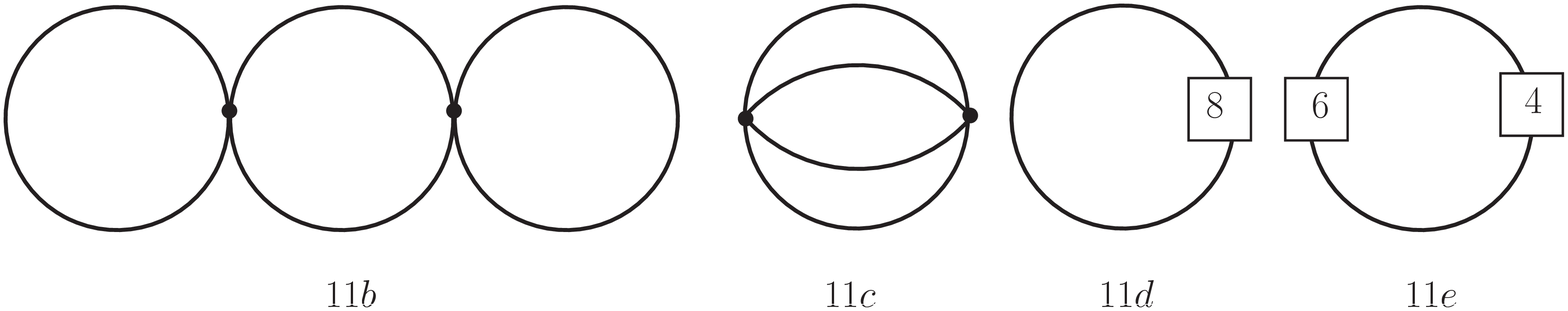}
\end{center}
\caption{\it Feynman graphs: Partition function of the $d_s$=3 ferromagnet up to three-loop order $p^{11}$ in the perturbative expansion.
Note that loop graphs are suppressed by three powers of momentum.}
\label{figure2}
\end{figure}

Let us proceed with ferromagnets in three space dimensions \cite{Hof02,Hof11a}. The relevant Feynman graphs for the partition function are
shown in Fig.~\ref{figure2}.
Loops are now suppressed by three powers of momentum, such that the structure of the temperature powers and the "communication" among the
graphs is rather different. In particular, the three-loop graphs (11a-c) are no longer of the same order as the two-loop graphs with
insertions from ${\cal L}^4_{eff}$ (10a-b). This seems to indicate a serious problem, as one might think that potential divergences in the 
three-loop graphs can no longer be absorbed into next-to-leading order effective constants of the two-loop graphs. However, as we have
pointed out, the cateye graph is not singular in three space dimensions. Also, there are no logarithmic terms in the temperature in these
series at three-loop order.

The leading contribution in the low-temperature expansion of the partition function is given by the one-loop graph 5 of order $p^5$,
giving rise to a $T^{5/2}$-term in the pressure. In contrast to QCD or ferromagnets in two space dimensions, we thus also have
half-integer powers of the temperature. The full three-loop results for the pressure and the spontaneous magnetization turn out to be
\cite{Hof11a}
\begin{equation}
\label{Pressure}
P = a_0 T^{\frac{5}{2}} + a_1 T^{\frac{7}{2}} + a_2 T^{\frac{9}{2}} + {\bf a_3 T^5} + {\bf a_4 T^{\frac{11}{2}}} + {\cal O}({\bf T^6}) ,
\end{equation}
and
\begin{equation}
\frac{\Sigma(T)}{\Sigma} = 1 - {\alpha}_0 T^{\frac{3}{2}} - {\alpha}_1 T^{\frac{5}{2}}
- {\alpha}_2 T^{\frac{7}{2}} - {\bf {\alpha}_3 T^4}
- {\bf {\alpha}_4 T^{\frac{9}{2}}} + {\cal O}({\bf T^5}) ,
\end{equation}
respectively. Again we have boldfaced all terms that are related to the Goldstone boson interaction. The coefficients in these series
depend on the low-energy effective constants. In the case of the spontaneous magnetization they are explicitly given by
\begin{eqnarray}
{\alpha}_0 & = & \frac{1}{8 {\pi}^{\frac{3}{2}} \Sigma {\gamma}^{\frac{3}{2}}} \,
\zeta(\mbox{$ \frac{3}{2}$}) , \nonumber \\
{\alpha}_1 & = & \frac{15 \, l_3}{16 {\pi}^{\frac{3}{2}} {\Sigma}^2
{\gamma}^{\frac{7}{2}}} \, \zeta(\mbox{$ \frac{5}{2}$}) \, , \nonumber \\
{\alpha}_2 & = & \frac{105}{32 {\pi}^{\frac{3}{2}} {\Sigma}^2 {\gamma}^{\frac{9}{2}}} \,
\Bigg( \frac{9 l^2_3}{2 \Sigma \gamma} \, - \, c_1 \Bigg) \,
\zeta(\mbox{$ \frac{7}{2}$}) , \nonumber \\
{\alpha}_3 & = & \frac{3 (8 l_1 + 6 l_2 + 5 l_3)}{64 {\pi}^3 {\Sigma}^3 {\gamma}^5}\
\zeta(\mbox{$ \frac{5}{2}$}) \, \zeta(\mbox{$ \frac{3}{2}$}) \, , \nonumber \\
{\alpha}_4 & = & \frac{945}{64 {\pi}^{\frac{3}{2}} {\Sigma}^2 {\gamma}^{\frac{11}{2}}} \,
\Bigg( d_1 - \frac{11 l_3 c_1}{\Sigma \gamma} \Bigg) \, \zeta(\mbox{$ \frac{9}{2}$})
- \frac{1}{2 \Sigma^3 \gamma^{\frac{9}{2}}} \, j_2 ,
\end{eqnarray}
where the constant $j_2$ stems from the cateye 11c graph and takes the value
\begin{equation}
j_2 = -0.00008 \, .
\end{equation}
Before the pioneering work by Dyson \cite{Dys56} and Zittartz \cite{Zit65}, it was quite unclear at which order in the low-$T$ series the
spin-wave interaction shows up in the spontaneous magnetization or in the other thermodynamic quantities. Here not only have we rederived
this result in a more elegant manner, but also have we extended the calculation up to three-loop order. Interestingly, the two-loop graphs
10a-b with insertions from the next-to-leading order Lagrangian ${\cal L}_{eff}^4$ represent the leading contribution from the spin-wave
interaction, while the three-loop graphs 11a-c only contribute at next-to-leading order.  Again, this is a conseqeuence of the
peculiar momentum counting in the nonrelativistic domain: in three space dimensions, loop are suppressed by three powers of momentum.

Let us have a closer look at the scales involved in QCD and the present case of the ferromagnet in three space dimensions. The series for
the respective order parameters, the quark condensate and the magnetization, are
\begin{equation}
\frac{\langle \bar{q} q \rangle(T,m_q)}{\langle 0 | \bar{q} q | 0 \rangle}
= 1 - {\beta}_0 T^2 - {\bf {\beta}_1 T^4} - {\bf {\beta}_2 T^6 \, ln T} + {\cal O}({\bf T^8}) ,
\end{equation}
and
\begin{equation}
\frac{\Sigma(T,H)}{\Sigma} = 1 - {\alpha}_0 T^{\frac{3}{2}} - {\alpha}_1 T^{\frac{5}{2}}
- {\alpha}_2 T^{\frac{7}{2}} - {\bf {\alpha}_3 T^4}
- {\bf {\alpha}_4 T^{\frac{9}{2}}} + {\cal O}({\bf T^5}) .
\end{equation}
In the chiral limit ($m_q \to 0$) and in zero magnetic field, the leading coefficients read
\begin{eqnarray}
{\beta}_0 & = & \frac{1}{8 F^2_{\pi}} , \nonumber \\
\alpha_0 & = & \frac{\zeta(\mbox{$ \frac{3}{2}$})}{8 {\pi}^{\frac{3}{2}} \Sigma {\gamma}^{\frac{3}{2}}} ,
\end{eqnarray}
such that the corresponding temperature scales differ in more than ten orders of magnitude,
\begin{equation}
\Lambda^T_{\mbox{\scriptsize{QCD}}} = \sqrt{8} F_{\pi} \approx 250 \mbox{\normalsize MeV} , \qquad
\Lambda^T_{\mbox{\scriptsize{F}}} = {\alpha}^{-2/3}_0 \approx 10 \mbox{\normalsize meV} .
\end{equation}
For temperatures small compared to $\Lambda^T_{\mbox{\scriptsize{QCD}}}$ and $\Lambda^T_{\mbox{\scriptsize{F}}}$, the low-temperature series are
perfectly valid. As we discuss below, the situation in one and two space dimensions is more subtle, because of the Mermin-Wagner theorem
\cite{MW68}.

\begin{figure}
\begin{center}
\includegraphics[width=16cm]{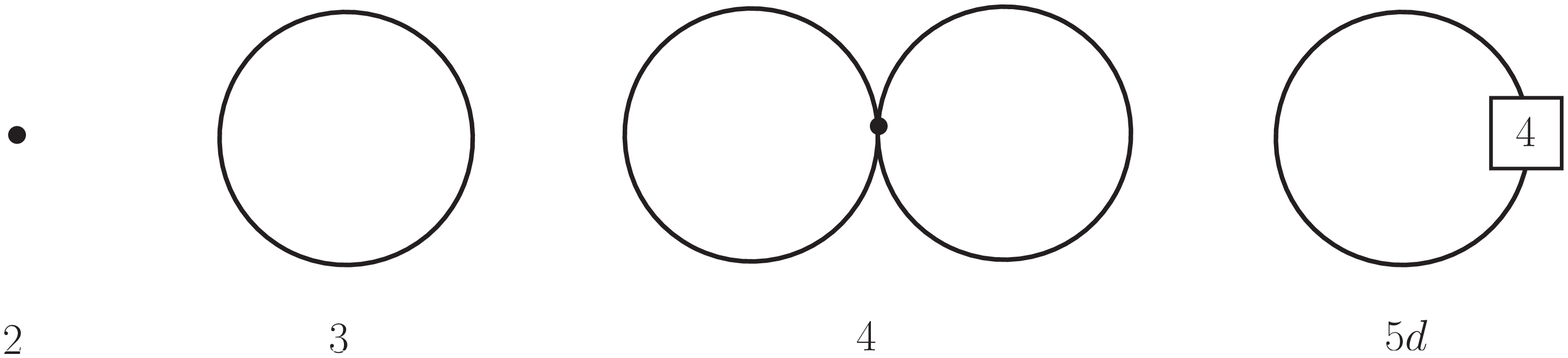}

\vspace{4mm}

\includegraphics[width=16cm]{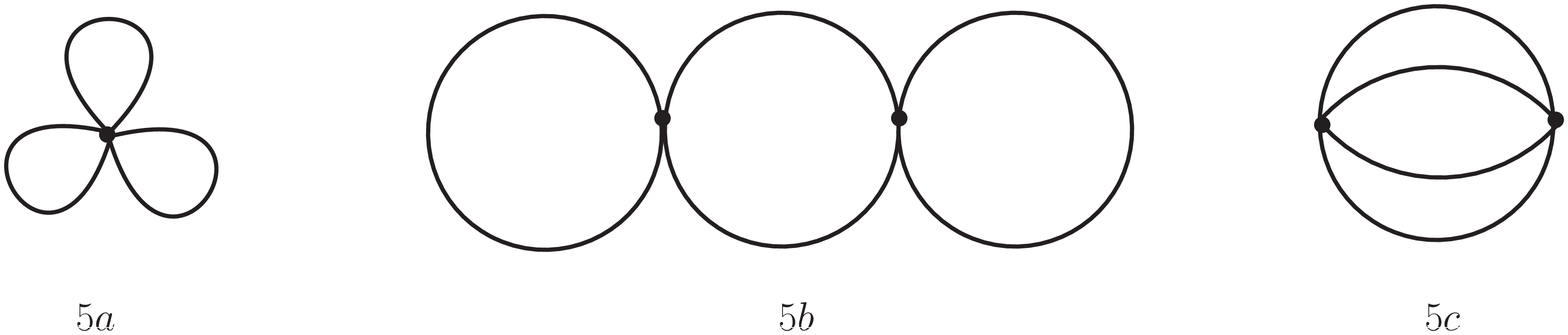}
\end{center}
\caption{\it Feynman graphs: Partition function of the $d_s$=1 ferromagnet up to three-loop order $p^5$ in the perturbative expansion.
Note that loop graphs are suppressed by one power of momentum.}
\label{figure3}
\end{figure}

Finally, we turn to ferromagnetic spin chains \cite{Hof13a,Hof13b}. Remember that in a Lorentz-invariant setting, the effective Lagrangian
method does not consistently work in one space dimension. However, ferromagnetic loops in one space dimension are still suppressed by one
power of momentum. The relevant Feynman graphs, again up to three-loop order in the perturbative expansion of the partition function, are
depicted in
Fig.~\ref{figure3}.
Remarkably, in one space dimension only very few effective constants are required. In particular, while ${\cal L}_{eff}^6$ and even
${\cal L}_{eff}^8$ contribute up to three-loop order in three and two space dimensions, in the case of ferromagnetic spin chains only
${\cal L}_{eff}^4$ is relevant -- as a matter of fact, it only occurs in the one-loop graph 5d, such that the spin-wave interaction is not
affected at all by subleading pieces of the effective Lagrangian.

For the pressure we obtain,
\begin{equation}
P = {\tilde a}_0 T^{\frac{3}{2}} + {\bf {\tilde a}_1 T^{\frac{5}{2}}} + {\cal O}({\bf T^3}) ,
\end{equation}
while the magnetization amounts to
\begin{equation}
\frac{\Sigma(T,H)}{\Sigma} = 1 - {\tilde \alpha}_0 T^{\frac{1}{2}} - {\bf {\tilde \alpha}_1 T^{\frac{3}{2}}} + {\cal O}({\bf T^2}) .
\end{equation}
Much like in the case of the ferromagnet in three space dimensions, there are no logarithmic terms in the temperature in these series, and
we also have half-integer powers of $T$. Note that the spin-wave interaction already sets in at next-to-leading order. The corresponding
coefficient ${\tilde a}_1$ turns out to be positive, such that the spin-wave interaction in the pressure is repulsive. Remarkably, this is
a rigorous statement following from the leading effective Lagrangian ${\cal L}_{eff}^2$ alone -- it is not affected by any higher-order
effective constants up to the three-loop level we have considered in the present analysis. 

At the end of this overview, we like to point out some subtleties that arise at finite temperature in space dimensions lower or equal 
than two. Although the spin symmetry O(3) of the underlying Heisenberg model is spontaneously broken at zero temperature, both for
ferromagnetic films and for ferromagnetic spin chains, matters are quite different at nonzero temperature. Following the Mermin-Wagner
theorem \cite{MW68}, there is no spontaneous symmetry breaking in space dimensions two and one in the isotropic Heisenberg model. Indeed,
the low-energy spectrum of both systems is characterized by a nonperturbatively generated energy gap. Accordingly, the correlation length
of magnons is no longer infinite (as for a true Goldstone bosons), but finite. In the case of ferromagnets in two dimensions, the
correlation length is still exponentially large \cite{KC89},
\begin{equation}
\xi_{np} = C_{\xi} a S^{-\frac{1}{2}} \, \sqrt{\frac{T}{J S^2}} \, \exp \! \Big[\frac{2 \pi J S^2}{T} \Big]  \qquad (d_s = 2) ,
\end{equation}
where $a$ is the distance between two neighboring sites on the square lattice, and $ C_{\xi} \approx 0.05$ is a dimensionless constant.
For ferromagnetic spin chains, on the other hand, the correlation length is only proportional to the inverse temperature \cite{Kop89},
\begin{equation}
\xi_{np} =  a \, C^{(0)}_{\xi} \, \frac{J S^2}{T} \, \Big[1 + C^{(1)}_{\xi} \frac{1}{\pi} \sqrt{\frac{T}{J S^3}} + {\cal O}(T)  \Big]
\qquad (d_s = 1) .
\end{equation}
Here $a$ is the distance between two neighboring sites of the spin chain and $C^{(0)}_{\xi}$ and $C^{(1)}_{\xi}$ are dimensionless constants. 

\begin{figure}
\begin{center}
\includegraphics[width=15cm]{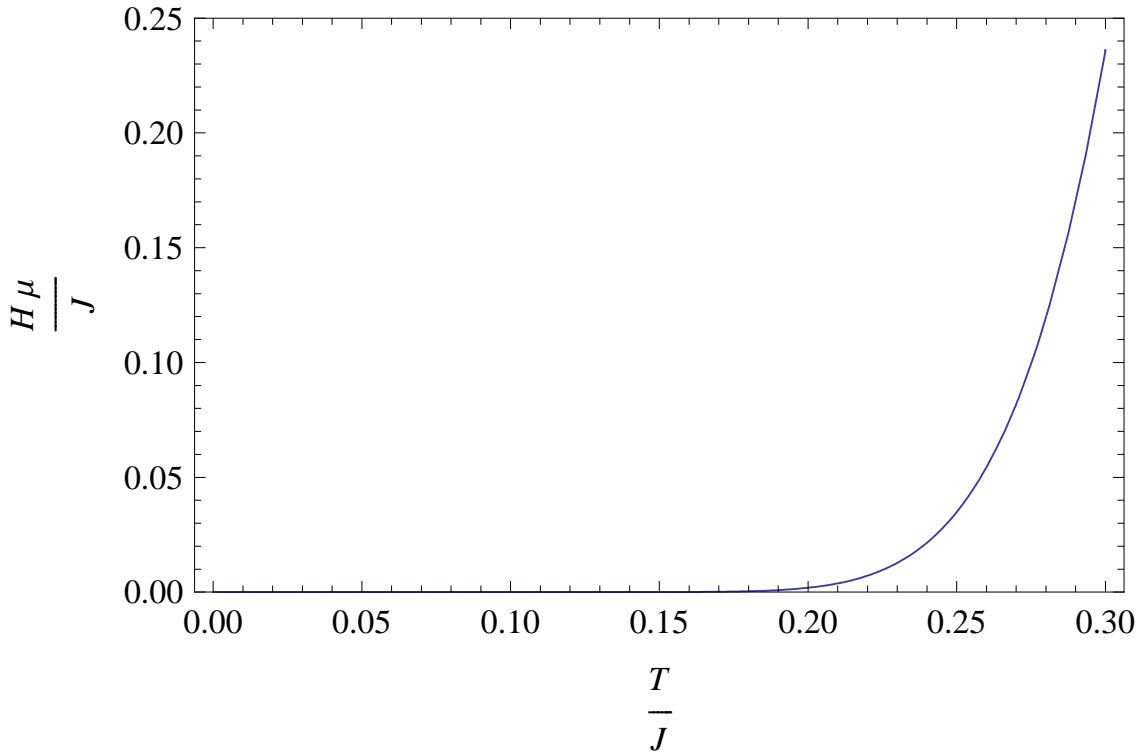}
\end{center}
\caption{\it Domain of validity for the ferromagnet in two space dimensions: in the parameter regime above the curve the effective series
are safe.}
\label{figure4}
\end{figure}

\begin{figure}
\begin{center}
\includegraphics[width=15cm]{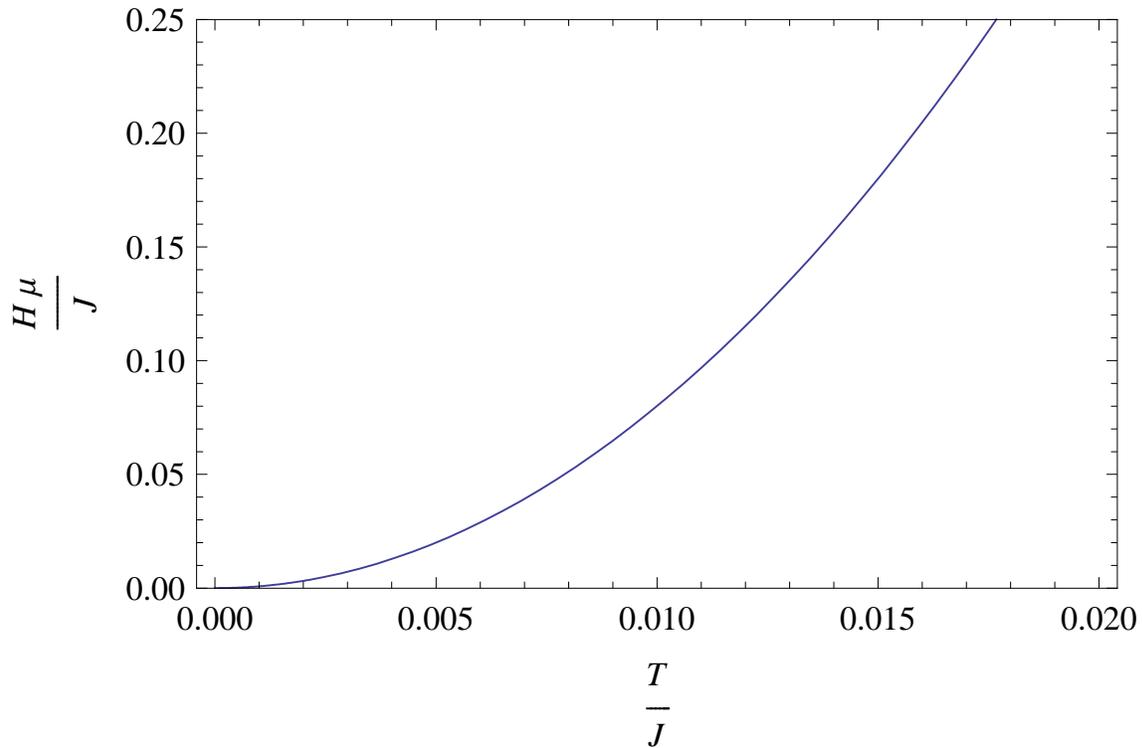}
\end{center}
\caption{\it Domain of validity for the ferromagnetic spin chain: in the parameter regime above the curve the effective series are safe.}
\label{figure5}
\end{figure}

As we have discussed in much detail in the original publications \cite{Hof12a,Hof12b,Hof13a,Hof13b}, this then implies that the external
magnetic field cannot be switched off completely, because we would then leave the validity domain of the effective low-temperature
expansion presented here. Still, the excluded regions are rather tiny, in particular in the case of ferromagnets in two space dimensions.
We have illustrated this in two plots: Fig.~\ref{figure4} refers to ferromagnetic films, while Fig.~\ref{figure5} corresponds to
ferromagnetic spin chains. In either case, in the parameter regime above the plotted curve, the effective expansion works perfectly. Note
however, that the horizontal axis lies  outside the allowed domain: the magnetic field cannot be switched off completely. This is
different from ferromagnets in three space dimensions or QCD: there, the chiral limit or the limit of zero external magnetic field does
not pose any problems: the effective series are valid as they stand.

\section{Conclusions}

The low-energy dynamics of systems with a spontaneously broken symmetry is dominated by the corresponding Goldstone bosons and a
systematic effective field theory can be constructed. While the effective field theory for QCD (chiral perturbation theory) has been
established a long time ago, the technique has been transferred and applied to nonrelativistic systems only recently. In the present
overview we have focused on the properties of QCD and ferromagnets in one, two and three space dimensions at finite temperature. In
particular, we have discussed the partition function and various observables derived from there, including order parameters, up to the
three-loop level.

The structure of these low-temperature series is an immediate consequence of the symmetries of the underlying theory, i.e., the QCD
Lagrangian and the Heisenberg Hamiltonian, respectively. Although ferromagnetic magnons obey a quadratic dispersion relation, there is an
interesting analogy between ferromagnets in two space dimensions and QCD: in either case, Goldstone boson  loops are suppressed by two
powers of momentum in the effective loop expansion. Accordingly, the next-to-leading order effective constants undergo logarithmic
renormalization and the ultraviolet divergences in three-loop graphs are absorbed into effective constants of ${\cal L}_{eff}^4$  arising
in two-loop graphs. Moreover, in the low-temperature expansion of the various thermodynamic quantities, terms logarithmic in the
temperature emerge at three-loop order, related to the Goldstone boson interaction.

On the other hand, ferromagnets in one or three space dimensions are rather different. In the latter case, loops are suppressed by three
powers of momentum, in the former case, loops are still suppressed by one momentum power. It is quite remarkable that the systematic
effective field theory method works in one space dimension in the nonrelativistic domain -- in a Lorentz-invariant setting, the method
does not work consistently in one space dimension, since loops are no longer suppressed. In the low-temperature series of the various
thermodynamic quantities, no logarithmic terms in the temperature show up in the case of ferromagnets in one and three space dimension.
Instead, we also have half-integer powers of the temperature. Furthermore, the lower the space dimension, the less higher-order pieces of
the effective Lagrangian and the less low-energy effective constants are needed up to the three-loop level.

The scales involved in QCD and in ferromagnets differ in about ten orders of magnitude. Still, the effective field theory captures the
low-temperature properties of both systems in a systematic way. What is important is not the absolute scale, but the question whether or
not the temperature is low compared to the specific scale involved.

It is important to point out that many papers have been devoted in the past to the problem of how the spin-wave interaction in
ferromagnets manifests itself in the partition function. In particular, in three space dimensions there are more than a hundred papers
available, that are based on different methods such as spin-wave theory, Schwinger boson mean field theory, Green's functions, etc. We
like to stress that many of these results are not only in contradiction with each other, but also in contradiction with the effective
field theory results presented here. The systematic effective field theory approach hence proves to be superior to conventional condensed
matter techniques, as it addresses the problem from a model-independent point of view.

\ack The author would like to thank U.-J. Wiese and H. Leutwyler for stimulating discussions.

\end{document}